\documentstyle[twocolumn,aps,epsf]{revtex}

\def\d{{\rm d}}

\def\vector#1{{\bf #1}}

\def\vk{{\vector k}}
\def\vq{{\vector q}}

\def\dps{\displaystyle}

\def\Tc{{T_{\rm c}}}
\def\Tcp{{T_{{\rm c}p}}}

\def\hightc{{high-$T_{\rm c}$ }}

\def\kB{{k_{\rm B}}}

\def\hsp#1{\hspace{#1ex}}

\def\Tc{{T_{\rm c}}}

\def\lsim{\stackrel{{\textstyle<}}{\raisebox{-.75ex}{$\sim$}}}
\def\gsim{\stackrel{{\textstyle>}}{\raisebox{-.75ex}{$\sim$}}}

\def\lesssim{\stackrel{{\textstyle<}}{\raisebox{-.75ex}{$\sim$}}}

\def\kF{k_{{\rm F}}}

\def\epsF{\epsilon_{\rm F}}

\def\omegaD{{\omega_{\rm D}}}

\def\UGe{{${\rm UGe_2}$}}

\begin{document}
\draft

\twocolumn[\hsize\textwidth\columnwidth\hsize\csname 
@twocolumnfalse\endcsname


\title{Triplet superconductivity induced by screened phonon \\ 
       interactions in ferromagnetic compounds} 


\author{Hiroshi Shimahara and Mahito Kohmoto$^{*}$} 


\def\runtitle{
Triplet superconductivity induced by screened phonon 
interactions}


\address{
Department of Quantum Matter Science, ADSM, Hiroshima University, 
Higashi-Hiroshima 739-8530, Japan \\
$^{*}$Institute of Solid State Physics, University of Tokyo, 
Kashiwa 277-8581, Japan
}





\date{Received ~~~ January 2001}

\maketitle

\begin{abstract}
We propose that screened pairing interactions mediated by phonons could 
give rise to a spin triplet superconductivity 
in ferromagnetic compounds such as \UGe. 
It is pointed out that the pairing interactions 
include anisotropic components such as those of $p$, $d$, $f$-waves 
in addition to dominant $s$-wave component 
due to the momentum dependence. 
Since the ferromagnetic long-range order coexists, 
there is a large splitting of the Fermi surfaces of up and down spin 
electrons, which suppresses singlet pairing. 
Therefore, triplet pairing occurs at last due to the sub-dominant 
anisotropic interactions, even in the absence of magnetic contribution 
to the pairing interactions. 
\end{abstract}

\pacs{
}


]

\narrowtext

Recently, a coexistence of superconductivity and ferromagnetic 
long-range order was observed in \UGe \hsp{0.25} under 
pressure~\cite{Sax00}. 
In the phase diagram on the pressure and temperature plane 
the superconductivity appears inside the area of the ferromagnetic phase. 
For this proximity of the superconductivity and the ferromagnetism, 
spin triplet superconductivity is a possible candidate in this compound.

In addition, singlet pairing is considered to be unfavalable 
in this compound. 
The ferromagnetism and superconductivity 
occur in the same electron band~\cite{Sax00,Shic00} or at least 
in very close electron bands from the crystal structure. 
Thus, there is a large Fermi surface splitting, 
which suppresses antiparallel spin pairing, 
in the electron band responsible to the superconductivity.

As a mechanism of triplet pairing, 
magnetically mediated superconductivity has been 
considered~\cite{Sax00}. 
However, there are some behaviors which are not easily explained 
only by this mechanism. 
For example, the magnetic fluctuations and their contribution to 
the pairing interactions increase near the transition points. 
Such behavior is reproduced in the calculation of 
Fay and Appel~\cite{Fay80}. 
In their calculation, 
it decreases in a narrow region of the width of 1\% of the exchange 
interaction parameter near the second order transition point, 
but this decrease would not occur in \UGe, 
since the magnetic transition is of first order in \UGe. 
Therefore, the superconducting transition temperature seems to 
increase as the magnetic boundary is approached in this mechanism. 
However, in the observation, $\Tc$ decreases 
near the phase boundary ($\sim 1.6{\rm GPa}$) 
within the width of $\sim 0.3{\rm GPa}$ of the pressure, 
which is not very narrow.

Decreases of the superconducting transition temperatures as the magnetic 
phase approached were also observed in \hightc \hsp{0.25} superconductors. 
However, in those compounds, the magnetic phase is antiferromagnetism. 
Thus, we have a physical explanation of the decrease of 
the superconducting transition temperature 
based on the reduction of the density of states 
near the Fermi surface (pseudogap) 
due to the antiferromagnetic fluctuations~\cite{Shi00a}. 
On the other hand, the ferromagnetic fluctuations do not induce 
any pseudogap near the Fermi surface, since it is split to two surfaces.

On the other hand, 
if the decrease of the superconducting transition temperature 
near the magnetic phase boundary at the high pressure is correlated 
to the decrease of the ferromagnetic transition temperature, 
the absence of the superconductivity at pressures 
$p \lesssim 1{\rm GPa}$ does not seem to be explained 
without any extra mechanisms. 
For example, pair breaking effect due to Lorentz force by internal 
magnetic field created by ferromagnetic moments might suppress the 
superconducting transition temperature at low pressure. 
It might be also possible that 
the degree of nesting of the Fermi surface~\cite{Shic00} 
changes by pressure and gives rise to the pressure dependences 
of $\Tc$ and $T_{x}$ defined in~\cite{Sax00}. 
Therefore, the magnetic mechanisms of the superconductivity is not 
conclusive at present.

In this paper, we propose a phonon mechanism of triplet superconductivity 
in \UGe \hsp{0.25} system. 
In the conventional mechanisms, the phonon mediated interactions 
have often been considered not to induce anisotropic superconductivity. 
However, we shall point out in this paper 
that the momentum dependence of the phonon mediated interactions 
could give rise to a triplet superconductivity 
in ferromagnetic superconductors. 
In a spin polarized state, the spin and charge degrees of freedom 
are locked together, and we concentrate ourselves on clarifying 
properties of the phonon mediated interaction.

It is easily verified that the pairing interactions contain both 
singlet and triplet components due to the momentum dependence. 
For example, when the screening length becomes longer for weak screening, 
the pairing interactions mediated by phonons have shaper peak near 
$q \sim 0$ in momentum space. 
Then, anisotropic components have large magnitudes. 
In the limit where the pairing interactions are proportional to 
the $\delta$-function, all components have the same magnitude. 
Foulkes and Gyorffy theoretically examined $p$-wave pairing in 
metals due to electron-phonon interaction, 
when the short range Coulomb interaction suppresses 
the $s$-wave pairing~\cite{Fou77}.

Anisotropic superconductivity induced by phonon interaction has been 
considered by Abrikosov as a mechanism of \hightc 
superconductivity~\cite{Abr94}. 
In this mechanism, it was shown that the gap function can vary 
in sign in the presence of on-site Coulomb repulsion. 
The momentum dependence of the pairing interactions was examined, 
where screening effect was taken into account. 
Bouvier and Bok calculated the superconducting gap and obtained 
anisotropic momentum dependence of the gap function~\cite{Bou95} 
in the same model. 
Recently, Friedel and Kohmoto~\cite{Fri00}, and 
Chang, Friedel, and Kohmoto~\cite{Cha00} have shown 
that $d$-wave superconductivity is induced by screened phonon 
interactions with an assist of a contribution from 
the antiferromagnetic fluctuations. 
In the studies so far, it was shown that a singlet pairing component 
is dominant in the screened phonon interactions.

Usually, since the largest component has even parity in the momentum space, 
it contributes to singlet pairing. 
Therefore, a sub-dominant triplet pairing interaction cannot induce 
superconducting transition. 
However, in ferromagnetic superconductors such as \UGe, 
the dominant singlet pairing interaction would not be able to overcome 
the Pauli pair breaking effect. 
Therefore, the sub-dominant interaction could give rise 
to the triplet superconductivity.

In the following, we illustrate that the phonon mediated pairing 
interactions contain both singlet and triplet pairing interactions 
by taking into account the screening effect. 
We calculate a coupling constant of sub-dominant triplet pairing interactions, 
which becomes dominant after singlet pairing is suppressed. 
It is shown that the screening effect gives rise to an additional lattice 
constant dependence of the pairing interactions 
through the screening length scaled by the inverse of the Fermi momentum.

We examine a model with a screened phonon interaction defined by 
\def\eqscrph{(1)}
$$
     V(\vk,\vk') = - g \frac{{q_s}^2}{q^2 + {q_s}^2} \cdot 
                \frac{{\omega(\vq)}^2}
                {{\omega(\vq)}^2 - (\xi_{\vk} - \xi_{\vk'})^2} , 
     \eqno\eqscrph
     $$
where $\vq = \vk - \vk'$, $q = |\vq|$, 
and ${q_s}^{-1}$ is the screening length. 
This interaction has been examined 
by many authors~\cite{Abr94,Bou95,Fri00,Cha00} 
for the \hightc superconductivity. 
As explained in the text books, near the Fermi surface where 
$|\xi| \lesssim \omega(\vq)$, the interaction is attractive due to 
over-screening. 
We put $(\xi_{\vk}-\xi_{\vk'})^2 \sim 0$ for simplicity, 
since $\kB T \ll \omegaD$ where $\omegaD$ denotes 
the Debye frequency. 
Hence, we have 
\def\eqscrphsimp{(2)}
$$
     V(\vk,\vk') 
     = - g \frac{{q_s}^2}{|\vk - \vk'|^2 + {q_s}^2} . 
     \eqno\eqscrphsimp
     $$

Let us first consider the Thomas-Fermi screening for a while 
for a qualitative argument. 
We will improve our theory with a more detailed momentum dependence 
of the dielectric funcion in the random phase approximation (RPA) 
later. 
In the Thomas-Fermi approximation, we put 
\def\eqscrqs{(3)}
$$
     {q_s}^2 = 4 \pi e^2 \rho_{\rm t}(\mu) , 
     \eqno\eqscrqs
     $$
which is valid for long wave lengths such as $q \ll \kF$. 
Here, $\rho_{\rm t}(\mu)$ is the total density of states of the conduction 
electrons per unit volume at the chemical potential $\mu$.

We assume a spherically symmetric Fermi surface for simplicity. 
Assuming $\omegaD \ll \epsF$, 
we can put $|\vk| \approx \kF$. Thus, we have 
\def\eqVinthetabar{(4)}
$$
     V(\vk,\vk') = - \frac{g \, (\alpha - 1)}
                   {\alpha - \cos{\bar \theta}} , 
     \eqno\eqVinthetabar
     $$
where ${\bar \theta}$ denotes the angle between $\vk$ and $\vk'$, 
and $\alpha = 1 + {q_s}^2/2\kF^2$.

The pairing interaction eq.~{\eqVinthetabar} is expanded as 
\def\eqVglgammagamma{(5)}
$$
     \begin{array}{rcl}
     V(\vk,\vk') 
     & = & \dps{
     \sum_{l = 0}^{\infty} V_{l} \, P_l(\cos{\bar \theta}) }\\
     & = & \dps{
     \sum_{l = 0}^{\infty} \sum_{m = -l}^{l}
     g_{l} \, \gamma_{lm}(\theta,\varphi) \gamma_{lm}^{*}(\theta',\varphi') ,
     }
     \end{array}
     \eqno\eqVglgammagamma
     $$
where $(\theta,\varphi)$ and $(\theta',\varphi')$ denote the directions 
of $\vk$ and $\vk'$, respectively. 
Here, 
$g_l = {V_l}/({2l + 1})$ and $
     \gamma_{lm}(\theta,\varphi) 
     = Y_{lm}(\theta,\varphi)$
with the spherical harmonic functions $Y_{lm}$.

The $s$-wave coupling constant is calculated as 
\def\eqgs{(6)}
$$
     \begin{array}{rcl}
     g_0 & = & V_0 
           = \dps{ \int V({\bar \theta}) 
                   \, \frac{\d {\bar \Omega}}{4 \pi} }\\[8pt]
         & = & \dps{ \frac{1}{2} 
                     g \, (\alpha-1) \log |\frac{\alpha-1}{\alpha+1}| . }
     \end{array}
     \eqno\eqgs
     $$
The coupling constant $g_0$ is larger than $g_1$, but it is not effective 
in practice, since singlet pairing is suppressed as we explained above. 
Thus, we calculate the next dominant $p$-wave component and obtain 
\def\eqVp{(7)}
$$
     \begin{array}{rcl}
     g_1 & = & \dps{ 
          \frac{1}{3} V_1 = \int V({\bar \theta}) \cos{\bar \theta} 
          \, \frac{\d {\bar \Omega}}{4 \pi}
          } \\[8pt] 
         & = & \dps{ 
          g \, (\alpha - 1) 
          {\Bigl [} 1 + \frac{\alpha}{2} 
          \log |\frac{\alpha-1}{\alpha+1}| {\Bigr ]} . 
          } 
     \end{array}
     \eqno\eqVp
     $$

In order to improve our approximation, 
we examine the screening effect in an RPA. 
The pairing interaction is modifed as 
\def\eqscrphLH{(8)}
$$
     \begin{array}{rcl}
     V(\vk,\vk') 
     & = & \dps{ 
     - g \frac{{q_s}^2}{q^2 + {q_s}^2 u(q/2\kF) } 
     \frac{{\omega(\vq)}^2}
                 {{\omega(\vq)}^2 - (\xi_{\vk} - \xi_{\vk'})^2}
     }
     \end{array}
     \eqno\eqscrphLH
     $$
with 
\def\equx{(9)}
$$
     u(x) = \frac{1}{2} {\Bigl [}
     1 + \frac{1-x^2}{2x} \log |\frac{1 + x}{1-x}| {\Bigr ]} , 
     \eqno\equx
     $$
according to the result of the dielectric function by Lindhard 
\def\eqkappaLindhard{(10)}
$$
     \kappa (q) = 1 + \frac{q_s^2}{q^2} \, u(\frac{q}{2\kF}) . 
     \eqno\eqkappaLindhard
     $$
The Thomas-Fermi approximation could be recovered 
in the long wave length limit. 
We apply the same simplification putting 
$(\xi_{\vk}-\xi_{\vk'})^2 \sim 0$ in eq.~{\eqscrphLH}. 
The coupling constant is obtained by numerical calculation.

\begin{figure}[htb]
\begin{center}
\leavevmode \epsfxsize=7cm  
\epsfbox{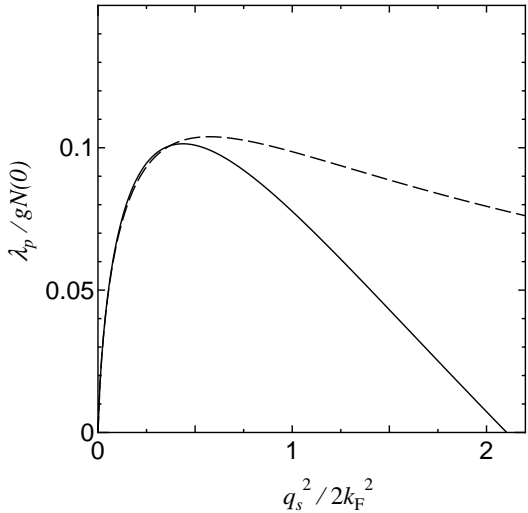}
\end{center}
\caption{Dimensionless coupling constant $\lambda_p$ of $p$-wave pairing 
in the unit of $gN(0)$ as a function of ${q_s}^2/2\kF^2$. 
Solid and dashed lines show results in the RPA and the Thomas-Fermi 
approximation, respectively. }
\label{fig:lambda}
\end{figure}

Figure~\ref{fig:lambda} shows the dimensionless coupling constants 
defined by 
$
     \lambda_p \equiv |g_1| N(0) , 
     $ 
where $N(0)$ is the density of states per a spin and unit volume, 
in the Thomas-Fermi approximation and an RPA. 
It is found that they exhibit a peak around 
${q_s}^2/2\kF^2 \sim 1/2$, 
and the Thomas-Fermi approximation agrees well with the RPA 
for ${q_s}^2/2\kF^2 \lsim 1/2$.

The transition temperature of the $p$-wave superconductivity is given by 
\def\eqTcren{(11)}
$$
     \Tcp = 1.13 \, \omegaD \exp[- 1/\lambda_p ] 
     \eqno\eqTcren
     $$
in the weak coupling limit. Short range Coulomb repulsion is 
not very effective in the case of $p$-wave pairing~\cite{Fou77}. 
Figure~\ref{fig:Tc} shows the transition temperature of 
the triplet superconductivity based on the formula {\eqTcren}. 
Because of the singular exponential form of eq.~{\eqTcren}, 
the peak of $\Tc$ is sharper than that of $\lambda_p$.

The screening length $q_s^{-1}$ is considered as follows. 
The total density of states per unit volume is expressed as 
\def\eqtotalDOS{(12)}
$$
     \rho_{\rm t}(\epsF) = 2 \frac{m}{2 \pi^2 \hbar^2} \kF . 
     \eqno\eqtotalDOS
     $$
Inserting eq.~{\eqtotalDOS} into eq.~{\eqscrqs}, we have 
\def\eqqsBohrradius{(13)}
$$
     \alpha - 1 = \frac{{q_s}^2}{2\kF^2} 
     = \frac{2}{\pi a_{\rm H}} \frac{1}{\kF} , 
     \eqno\eqqsBohrradius
     $$
where $a_{\rm H}$ denotes the Bohr radius 
($a_{\rm H} = \hbar^2/e^2 m \approx 0.5293{\rm \AA}$). 
Since $\kF \propto 1/a$ with lattice constant $a$, 
this expression means that 
a large lattice constant results in 
a strong screening effect, (i.e., a short screening length ${q_s}^{-1}$).

\begin{figure}[htb]
\begin{center}
\leavevmode \epsfxsize=7cm  
\epsfbox{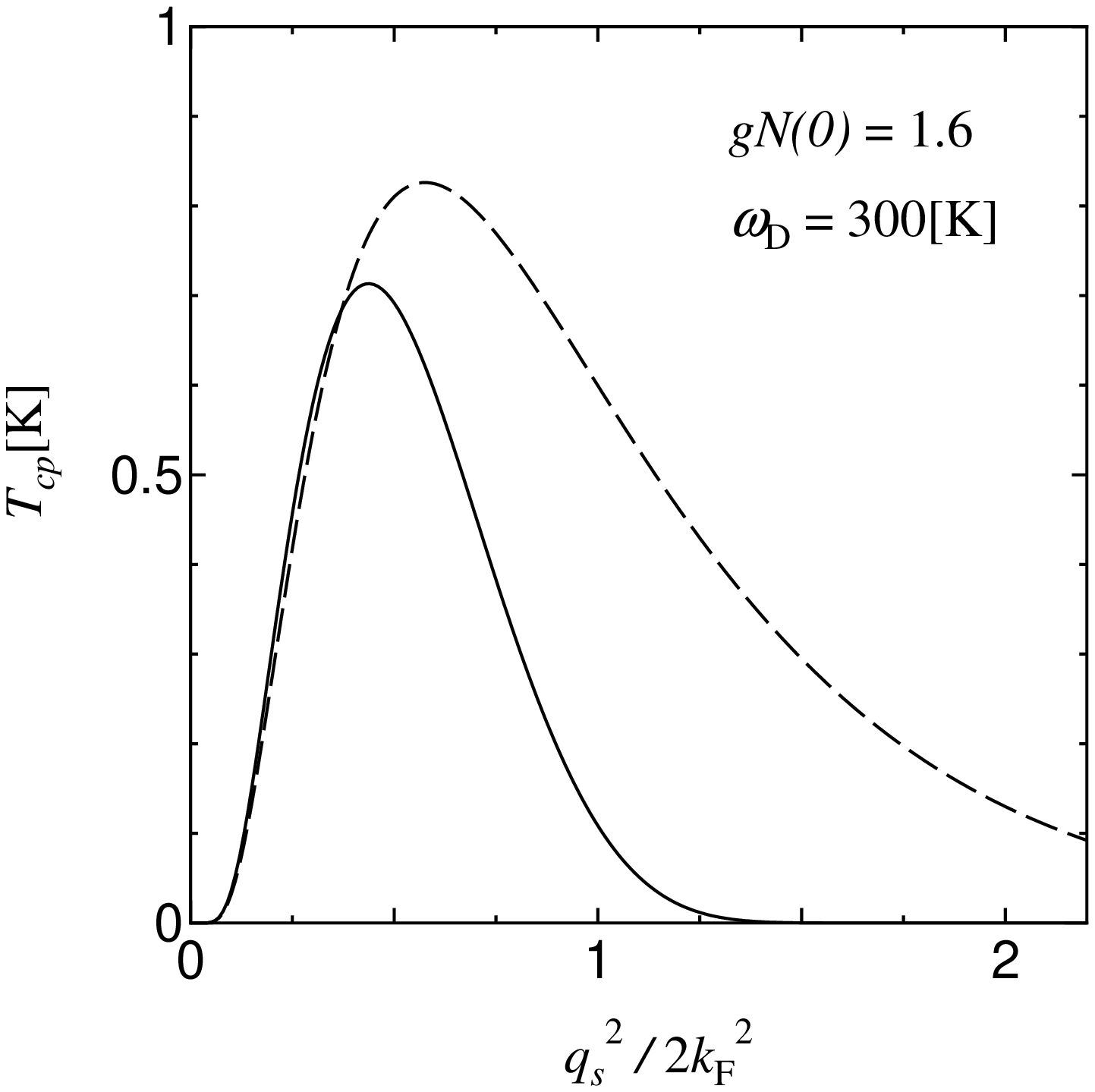}
\end{center}
\caption{
Superconducting transition temperature $\Tcp$ of $p$-wave pairing 
as a function of ${q_s}^2/2\kF^2$ 
calculated by eq.~{\eqTcren}. 
$\omegaD = 300{\rm [K]}$ and $gN(0) = 1.6$ are assumed as an example. 
Solid and dashed lines show results in the RPA and the Thomas-Fermi 
approximation, respectively. }
\label{fig:Tc}
\end{figure}

In our spherically symmetric system, the Fermi momentum $\kF$ is 
expressed as $\kF = (3 \pi^2)^{1/3}n^{1/3}/a$, 
where $n$ denotes conduction electron (or hole) number per a site. 
Thus, eq.~{\eqqsBohrradius} is written as 
\def\eqqsamn{(14)}
$$
     \frac{{q_s}^2}{2\kF^2} 
     = \frac{2}{\pi (3 \pi^2)^{1/3} n^{1/3}}
       \frac{a}{a_{\rm H}} 
     \equiv \frac{a}{a_0} . 
     \eqno\eqqsamn
     $$
Here, we have defined 
a characteristic length $a_0$ by 
\def\eqazerodef{(15)}
$$
     a_0 = \frac{1}{2} \pi (3 \pi^2)^{1/3} n^{1/3} a_{\rm H} , 
     \eqno\eqazerodef
     $$
which scales the lattice constant. 
If we put $n \sim 1$ as the order of magnitude, 
we obtain $a_0 = 2.57{\rm \AA}$. 
When $a \sim 4{\rm \AA}$~\cite{Mak59,Oik96}, 
one has ${q_s}^2/2\kF^2 = a/a_0 \sim 1.6$ as the order of magnitude.

Now, we discuss the pressure dependence of the transition temperature 
of the triplet supercoductivity. 
The pressure dependence of the transition temperature is not easily 
figured out theoretically in the phonon mechanisms 
even in $s$-wave superconductors. 
The Debye frequecy $\omegaD$ increases with pressure. 
Then, the prefactor of the weak coupling expression of $\Tc$ increases. 
On the other hand, the electron-phonon coupling constant and 
the electron density of states would decrease. 
These effects would contribute to the the pressure dependence 
in conventional superconductors. 
In \UGe, however, the superconducting transition temperature 
increases more sensitively to the pressure near 1GPa~\cite{Sax00} 
than those in conventional superconductors.

\begin{figure}[htb]
\begin{center}
\leavevmode \epsfxsize=7cm  
\epsfbox{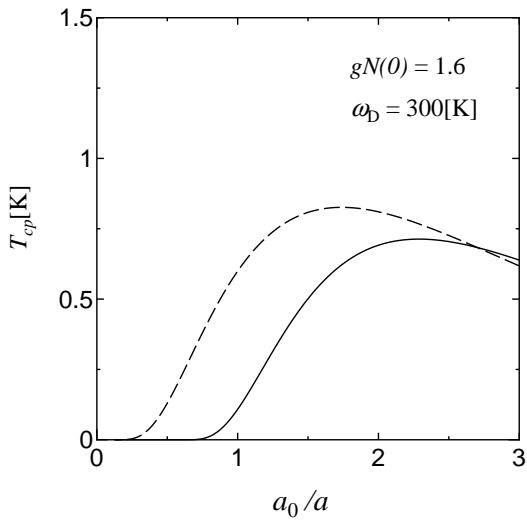}
\end{center}
\caption{
Superconducting transition temperature $\Tcp$ of $p$-wave pairing 
as a function of $a_0/a$ calculated by eq.~{\eqTcren}. 
$\omegaD = 300{\rm [K]}$ and $gN(0) = 1.6$ are assumed as an example. 
Solid and dashed lines show results in the RPA and the Thomas-Fermi 
approximation, respectively. }
\label{fig:Tcadep}
\end{figure}

In the present mechanism, we have an additional contribution to the 
pressure dependence through the screening effect. 
At high pressures, the lattice constant $a$ is shorten. 
Then, from eq.~{\eqqsamn}, the value of ${q_s}^2/2\kF^2$ is reduced 
by the pressure. 
Figure~\ref{fig:Tcadep} shows dependence of the superconducting 
transition temperature on the inverse of the lattice constant. 
A large value of $1/a$ corresponds to a high pressure. 
Since the pressure dependences of $\omegaD$, $g$, $N(0)$ are not 
considered, this figure only shows the additional contribution though 
the screening effect. 
It is found that the superconductivity does not occur for $a \gsim a_0$ 
and that the pressure would enhance the transition temperature 
of the triplet superconductivity near $a \sim a_0$. 
This result is consistent with the behavior of the transition temperature 
in the experimental phase diagram near 1GPa qualitatively, 
although it is more sensitive to the pressure in the experiment.

In addition, inside the region of the ferromagnetic phase, 
Lorentz force created by internal magnetic field should 
suppress pair formation. 
The superconducting transition temperature should be reduced from 
the value that we estimated in the above. 
Thus, the superconductivity must be absent in a wider region 
than those in Fig.~\ref{fig:Tc} and Fig.~\ref{fig:Tcadep}.

On the other hand, decreases of the superconducting and the ferromagnetic 
transition temperatures are observed at high pressures 
($\sim 1.6{\rm GPa}$)~\cite{Sax00}. 
For such high pressures, crystal structure might become unstable. 
Single phases might not be able to exist in this region. 
The decreases of the transition temperatures might be due to 
the instability of the crystal. 
At very high pressures, an unharmonicity may play some role in the 
pressure dependence of the superconducting transition temperature.

In addition, near the phase boundary between the ferromagnetic and 
paramagnetic states, the ferromagnetic fluctuations should be strong. 
The superconductivity might be suppressed near the phase boundary 
by some renormalization effect due to the strong magnetic fluctuations, 
since the pairing interactions would not increase significantly there 
in phonon mechanisms. 
In a spin polarized state the spin and charge degrees of freedom are 
locked together. The pressure dependence of $\Tc$ and $T_{x}$ might 
be explained by taking into account the change in the degree of 
nesting of the quasi-two-dimensional parts of 
the Fermi surface~\cite{Shic00}.

In conclusion, we proposed a phonon mechanism of triplet superconductivity 
in ferromagnetic systems. 
The momentum dependence of the pairing interactions mediated by phonons 
give rise to both singlet and triplet components. 
The singlet interaction is stronger than the triplet one, but the 
Fermi surface splitting suppresses the singlet superconductivity 
in the ferromagnetic state. 
For an appropriate region of the screening length, 
the triplet superconductivity occurs. 
The superconducting transition temperature depends on the lattice constant 
through the screening length scaled by inverse of the Fermi momentum. 
Such dependence might contribute to the pressure dependence of 
the superconducting transition temperature in \UGe.

This work was partially (H.S.) supported by a grant for Core Research 
for Evolutionary Science and Technology (CREST) from Japan Science and 
Technology Corporation (JST).


\end{document}